\documentclass[12pt]{nostro}
\usepackage{epsfig}\parskip 5pt plus 1pt
\newcommand{\be}{\begin{equation}}
\newcommand{\ee}{\end{equation}}
\newcommand{\bea}{\begin{eqnarray}}
\newcommand{\eea}{\end{eqnarray}}

\def\simge{\mathrel{%
   \rlap{\raise 0.511ex \hbox{$>$}}{\lower 0.511ex \hbox{$\sim$}}}}
\def\simle{\mathrel{
   \rlap{\raise 0.511ex \hbox{$<$}}{\lower 0.511ex \hbox{$\sim$}}}}

\begin{document}
\begin{frontmatter}
\thispagestyle{empty}
\begin{flushright}
{IFT-UAM/CSIC-10-31}\\
{IFIC-UV/CSIC-10-14}\\
{EURONU-WP6-10-17}
\end{flushright}
\title{The $\tau$-contamination of the golden muon sample at the Neutrino Factory} 
\author[Madrid,Valencia]{A. Donini}
\author[Valencia]{J.J. G\'omez Cadenas}
\author[Wurzburg]{D. Meloni}
\address[Madrid]{ I.F.T., Universidad Aut\'onoma de Madrid/CSIC, 28049 Madrid, Spain}
\address[Valencia]{I.F.I.C., Universitat de Valencia/CSIC, 46071 Valencia, Spain}
\address[Wurzburg]{Institut f{\"u}r Theoretische Physik und Astrophysik, 
Universit{\"a}t W{\"u}rzburg, D-97074 W{\"u}rzburg, Germany} 
\vspace{.2cm}
\begin{abstract}
We study the contribution of $\nu_e \to \nu_\tau \to \tau \to \mu$~transitions to 
the wrong-sign muon sample of the {\em golden} channel of the Neutrino Factory. Muons from tau decays are not really a background, since they contain information from the oscillation signal, and represent a small fraction of the sample. However, if not properly handled they introduce serious systematic error, in particular if the detector/analysis are sensitive to muons of low energy. This systematic effect is particularly troublesome for large $\theta_{13} \geq 1^\circ$ and prevents the use of the Neutrino Factory as a precision facility for large $\theta_{13}$. Such a systematic error disappears if the tau contribution to the golden muon sample is taken into account. The fact that the fluxes of the Neutrino Factory are exactly calculable permits the knowledge of the tau sample due to the $\nu_e \to \nu_\tau$~oscillation. We then compute the contribution to the muon sample arising from this sample {\em in terms of the apparent muon energy}. This requires the computation of a migration matrix $M_{ij}$~which describes the contributions of the tau neutrinos of a given energy $E_i$, to the muon neutrinos of an apparent energy  $E_j$. We demonstrate that applying $M_{ij}$~ to the data permits the full correction of the otherwise intolerable systematic error.    
\end{abstract}

\vspace*{\stretch{2}}
\begin{flushleft}
  \vskip 2cm
  \small
{PACS: 14.60.Pq, 14.60.Lm  }
\end{flushleft}
\end{frontmatter}

%
%
\newpage
%
%
\section{Introduction}
\label{sec:intro}

It has been understood for a long time that muons are the best experimental tool to study the yet--unknown parameters of the PMNS matrix \cite{neutrino_osc} 
(the small angle $\theta_{13}$~and the Dirac--CP violating phase $\delta$) at the Neutrino Factory. 

In a Neutrino Factory \cite{geer,derujula}, two intense beams of neutrinos are produced by the decay of muons which have previously been accelerated to high energy and stored in a ring. The so-called golden signature  \cite{Cervera:2000kp} is due to the oscillation $\nu_e \to \nu_\mu$, followed by the interaction
$\nu_\mu \to \mu$ in a detector. The resulting muon has the opposite sign of the lepton circulating in the storage ring, thus the name ``wrong--sign'' muon \cite{geer}.

The conventional detector considered at the Neutrino Factory is a large iron calorimeter, the so--called Magnetized Iron Detector (MIND) proposed in Ref.~\cite{Cervera:2000vy} and whose detailed performance was first studied in Ref.~\cite{Cervera:2000kp}. Those original studies focused on achieving a very pure sample of wrong sign muons. To that extent, stringent cuts in the muon momentum and transverse momentum with respect to the hadronic jet ($Q_t$) were introduced. 
As a result, the selection efficiency for neutrinos of energies less than 10 GeV was very low. 

It was first understood in Ref.~\cite{Burguet-Castell:2001ez} that eliminating low-energy neutrino bins from the oscillation analysis introduced serious problems in the joint 
determination of $\theta_{13}$~and $\delta$, namely the appearance of correlations and degeneracies (CAD) \cite{Burguet-Castell:2001ez,minadeg,th23octant,eight}. 
Understanding those CAD has been the subject of much theoretical work, both in the framework of Neutrino Factories (see, for example, Refs.~\cite{mina,Donini:2002rm,Burguet-Castell:2002qx,Donini:2003vz,disapnufact,Huber:2006wb} and references therein) or at other facilities such as 
Beta-Beams \cite{zucchelli,Burguet-Castell:2003vv,disappear,Burguet-Castell:2005va,pee,Huber:2005jk,Coloma:2007nn} or Super-Beams \cite{t2ksimulation,Minakata:2004pg,cernmemphys}. 

Although an iron calorimeter appears limited when it comes to detecting low energy muons, refined recent analyses \cite{ISS_det} have shown a much better neutrino efficiency than the original studies. The ability to include low-energy neutrino bins mitigates 
greatly the CAD problem at the Neutrino Factory \cite{ISS_phys}.

The wrong-sign muon sample is contaminated by backgrounds other than the oscillation signals. Those backgrounds are reduced by experimental cuts, although a high-efficiency analysis for low energy muons must tolerate a larger background, coming primarily from neutral currents in which a hadron manages to fake a low energy muon. Those backgrounds, however, introduce essentially a ``white noise'' to the signal (they contain no information of the oscillation being studied) and can be statistically subtracted. 

However, the oscillation  $\nu_e \to \nu_\tau$~introduces a small but significant sample of
``bona--fide'' wrong--sign muons, due to the subsequent production of a wrong--sign tau which decays in turn in a muon of the same sign. Notice that the wrong sign taus are not background but as good signal as the wrong sign muons, but the resulting muons, if not properly treated, introduce a systematic error in the muon sample. 

There are two reasons why this systematic error is troublesome:
\begin{enumerate}
\item In the Neutrino Factory, the total neutrino energy is computed by adding the energy of the muon to the energy of the hadronic jet. This operation yields a wrong result when the muon comes from a tau decay and it is detected at MIND, since no additional information regarding the neutrino missing energy in the decay can be provided (in contrast to the case of ECC or Liquid Argon detectors);
\item Muons from tau decay tend to accumulate in low-energy muon bins, since the missing neutrinos in tau decay result in a ``secondary'' muon which has, on average 1/3 of the tau energy.
\end{enumerate}

On the other hand, the joint determination of $\theta_{13}$~and $\delta$~is particularly sensitive to the low--energy muon neutrinos. The small contribution of the taus is in fact significant at
low energy, in particular for large $\theta_{13}$. What is worse is that it is not white noise, since it carries oscillation information different from that of the $\nu_e \to \nu_\mu$ golden channel. 
Notice that the muons from the tau sample pollute all measurements that use muons in the final state as the signal sample. This problem was already discussed in the context of precision measurements of the atmospheric parameters $(\Delta m_{32}^2,\theta_{23})$ at a Neutrino Factory in Ref.~\cite{indians}. In that paper, 
the authors were interested in the study of  the $\nu_\mu$ disappearance transition $\nu_\mu \to \nu_\mu$ at the Neutrino Factory, with the signal represented by the right-sign muon
sample in the INO detector. The problem arises, in this case, from $\tau$'s generated through the leading oscillation channel $\nu_\mu \to \nu_\tau$, with 
subsequent CC interactions $\nu_\tau N$. These $\tau$'s will eventually decay (with a $\sim$ 20\% branching ratio)  into muons, thus contaminating the right-sign muons sample, 
with a resulting precision loss in the $(\Delta m_{32}^2,\theta_{23})$ measurement. 

Trying to eliminate muons from wrong- or right-sign taus by using kinematical criteria is not a sound procedure, since any cuts which suppress them sufficiently will also deplete the low energy muon neutrino bins. Nor is it necessary, since the problem can be correctly treated by using migration matrices, a technique already introduced in Ref.~\cite{Burguet-Castell:2003vv} to correct the effect of computing the neutrino energy from the observed lepton in quasielastic events (QE).  In this paper, we estimate the systematic error introduced in the joint measurement of $\theta_{13}$ and $\delta$ if the effect is not treated properly and demonstrate that it ruins the sensitivity of the neutrino factory for large $\theta_{13}$. Next we show that such systematic error disappears if the tau contribution to the golden muon sample is taken into account. The fact that the fluxes of the Neutrino Factory are exactly calculable permits the knowledge of the tau sample due to the $\nu_e \to \nu_\tau$~oscillation. We then compute the contribution to the muon sample arising from this sample {\em in terms of the apparent muon energy}. This requires the introduction of a migration matrix $M_{ij}$~which describes the contributions of the tau neutrinos of a given energy $E_i$, to the muon neutrinos of an apparent energy  $E_j$. We demonstrate that applying $M_{ij}$~ to the data permits the full correction of the otherwise intolerable systematic error.    

\section{The source of the $\tau$-contamination problem}
\label{sec:taucont}
 
\begin{figure}[t!]
\vspace{-0.8cm}
\begin{center}
\begin{tabular}{cc}
\hspace{-0.4cm} \epsfxsize6cm\epsffile{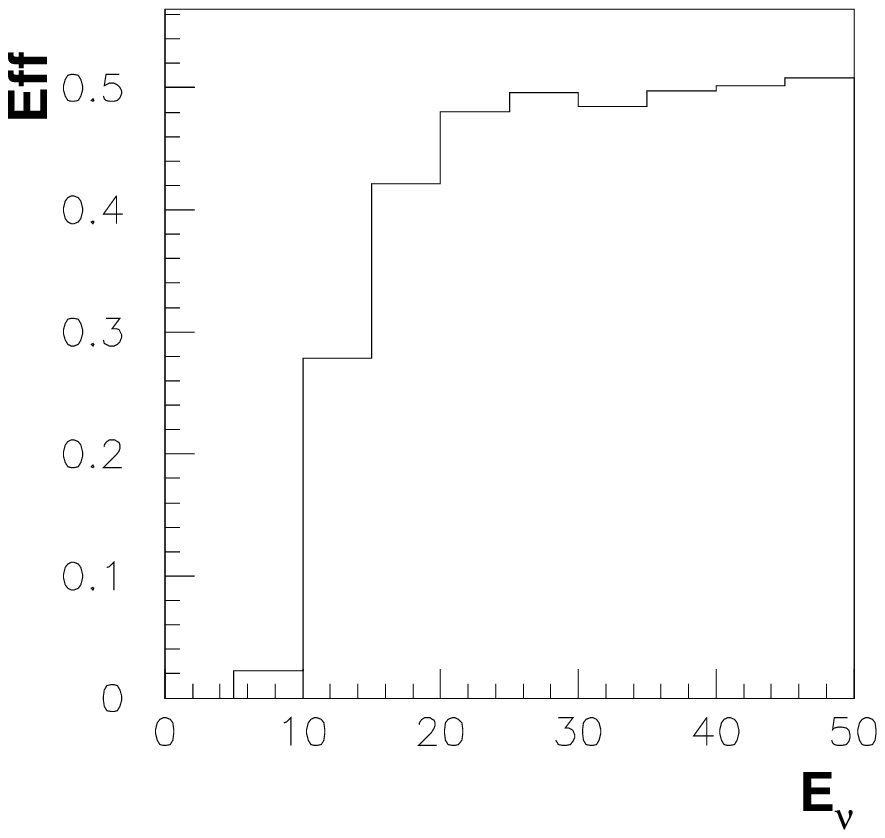} &
\hspace{-0.3cm} \epsfxsize6.5cm\epsffile{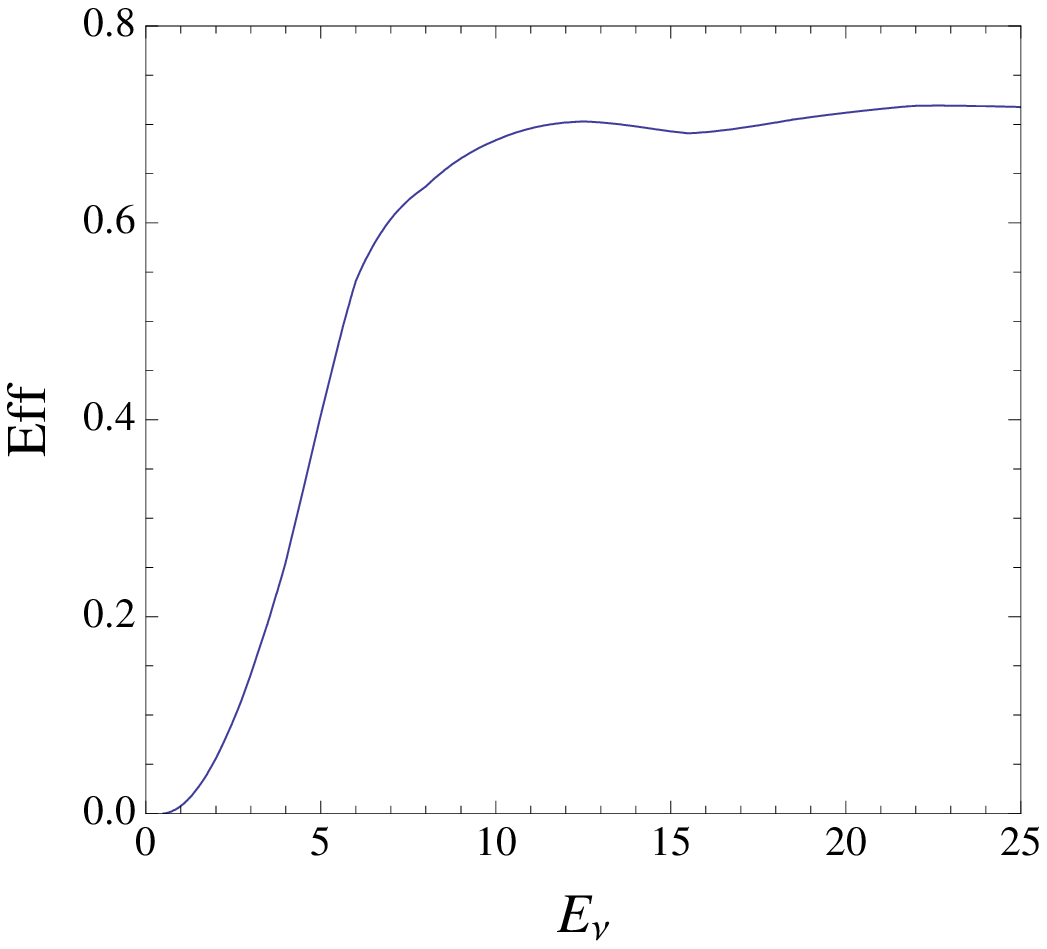}
\end{tabular}
\caption{\it Wrong-sign $\mu^-$ identification efficiency at the Neutrino Factory as a function of the reconstructed neutrino energy from Ref.~\cite{Cervera:2000kp} (left) and Ref.~\cite{ISS_det} (right).}
\label{fig:efficiencies}
\end{center}
\end{figure}

Figure~\ref{fig:efficiencies}, shows the muon neutrino reconstruction efficiency in the original MIND analysis 
from Ref.\cite{Cervera:2000kp} (left) and for the improved analysis discussed in Ref.~\cite{ISS_det} (right). Notice that in the early analysis the 
efficiency below 10 GeV is practically zero. All the subsequent studies who used this efficiency curve had, thus, an effective low-energy threshold around 10 GeV. 

However, all the studies on the degeneracy problem at the Neutrino Factory have shown that the signal content of these low-energy bins is extremely important to solve, or at least considerably mitigate the CAD problem: some of the degeneracies are strongly energy-dependent, and non-vanishing signal for all ranges of neutrino energies are required to distinguish
the true solution from its "clones" \cite{Donini:2003vz}. For this reason recent studies have focused in achieving better neutrino efficiency at the cost of accepting higher backgrounds. In Figure~\ref{fig:efficiencies} (right), the efficiency for events with reconstructed neutrino energy in the energy range $E_\nu \in [5,10]$ GeV is around 60\% and, even below 5 GeV, a non-negligible efficiency is found. 

The improved efficiency in the low-energy part of the neutrino spectrum, however, has as the drawback that a previously irrelevant background becomes now potentially harmful. 
The problem arises from the contamination of the wrong-sign muon sample by wrong-sign muons produced in the decay of 
wrong-sign $\tau$'s. It is well known \cite{Cervera:2000kp,Donini:2002rm} that $\nu_e$ oscillate into $\nu_\mu$ (the "golden" channel) and $\nu_\tau$ (the "silver" channel) with similar rates for the neutrino energies and baselines considered in a standard multi-GeV Neutrino Factory setup. Oscillations of $\nu_e$ into $\nu_\tau$ will produce $\tau$'s through $\nu_\tau N$ CC interactions within the detector. 
Approximately 20\% of the $\tau$'s will, eventually, decay into muons. Notice that these wrong-sign muons from wrong-sign $\tau$'s escape essentially all filters designed to kill the dominant backgrounds and directly add to the wrong-sign muon sample. 

Certainly if the events were to be divided in muon energy bins (rather than in bins of neutrino energy), muons arising from tau decay could be directly treated as an additional source of signal. However, the total energy of the neutrino is a fundamental input to separate high--energy charged currents from the low--energy dominant neutral current background. Thus, in the standard MIND analysis, the neutrino energy is induced by adding the energy of the muon and that of the hadronic jet. 

This approach, however, implies that the sample of wrong-sign muons from the decay of wrong-sign taus
will be distributed erroneously in neutrino energy bins, since the missing energy in the $\tau \to \nu_\tau \bar \nu_\mu \mu^-$ decay cannot be measured\footnote{Notice that this situation is different from what happens in an Emulsion Cloud Chamber, where the identification of the $\tau$ decay vertex allows for a separation of "golden" from "silver" muon samples \cite{Donini:2002rm}.}. 
The muons from tau decay will result, therefore, in a contamination of the wrong-sign muon sample by events whose parent neutrino energy is reconstructed wrongly: this is what has been called the 
"$\tau$-contamination" problem in \cite{indians}. 

\begin{figure}[t!]
\vspace{-0.5cm}
\begin{center}
\hspace{-0.3cm} \epsfxsize15cm\epsffile{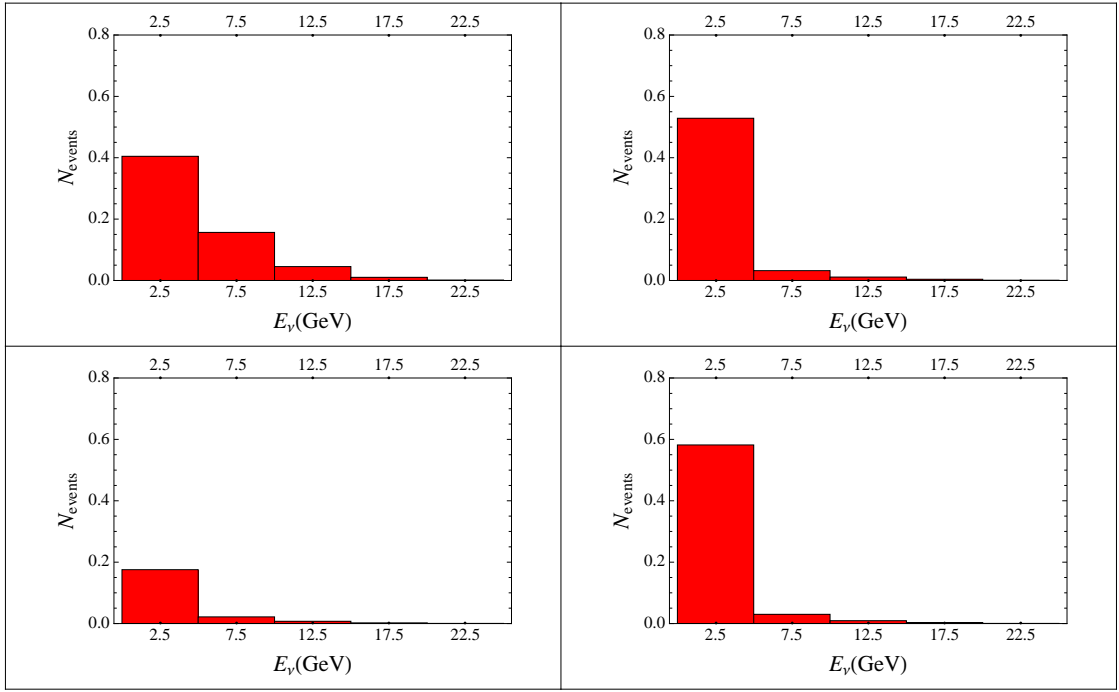} 
\caption{\it 
The fraction of $\tau$ contamination of the golden muon sample as a function of the reconstructed neutrino energy, for $\theta_{13} = 2^\circ$. 
Left: $L = 4000$ Km; Right $L = 7500$ Km.
Top: $\delta = -90^\circ$; Bottom: $\delta = +90^\circ$. 
}
\label{fig:taupercent}
\end{center}
\end{figure}

In Fig.~\ref{fig:taupercent} we show the fraction of muons coming from $\tau$-decay that can be found in the wrong-sign muon sample after binning in the 
reconstructed neutrino energy, using the new MIND efficiency, for a 25 GeV Neutrino Factory with detectors located at $L = 4000$ Km and at the so--called magic baseline, $L = 7500$ Km
(see Ref.~\cite{ISS_phys} for a discussion of why those are the two optimal baselines). 
The data are shown for $\theta_{13} = 2^\circ$ and two values of $\delta$, $\delta = \pm 90^\circ$.
In the left panels, we show the level of $\tau$-contamination at the $L = 4000$ Km baseline and in the right, at $L = 7500$ Km. 

In all cases, a significant $\tau$-contamination 
can be observed below 5 GeV. At the shorter baseline---which is optimized to increase sensitivity to $\delta$---, the amount of $\tau$-contamination is $\delta$-dependent: 
we find that 40\% (20\%) of the muons are produced through $\tau$-decay for $\delta = -90^\circ (90^\circ)$. Not so at the long baseline which has been chosen to be largely insensitive to 
$\delta$. Here we find that $\tau$-contamination below 5 GeV is  about 60\%, independently of the value of $\delta$. With the exception of the top left panel 
(corresponding to $\theta_{13} = 2^\circ, \delta = -90^\circ, L = 4000$ Km), the $\tau$-contamination in the energy range $E_\nu \in [5,10]$ GeV is at the percent level. 
Above 10 GeV, the $\tau$-contamination becomes negligible (for a 25 GeV Neutrino Factory). For this reason, in previous studies of the Neutrino Factory
performances obtained using the original MIND efficiency (i.e., the muon identification efficiency of Fig.~\ref{fig:efficiencies}, left), the $\tau$-contamination problem
was absent in practice.

Since the far baseline contributes to data with a substantially $\delta$-blind component (both for the true golden channel data and for the corresponding $\tau$-contamination), 
it is not able to remove the distortion induced by  a wrong treatment of muons from wrong-sign taus. For this reason, in the next section we will focus on the intermediate baseline, 
$L = 4000$ Km, and show only at the end of the section that the combination of the two baselines does not solve the problem.

%
\section{The impact of the $\tau$-contamination problem}
\label{sec:pr}
%
%

\begin{figure}[t!]
\vspace{-0.5cm}
\begin{center}
\begin{tabular}{cc}
\hspace{-0.3cm} \epsfxsize6cm\epsffile{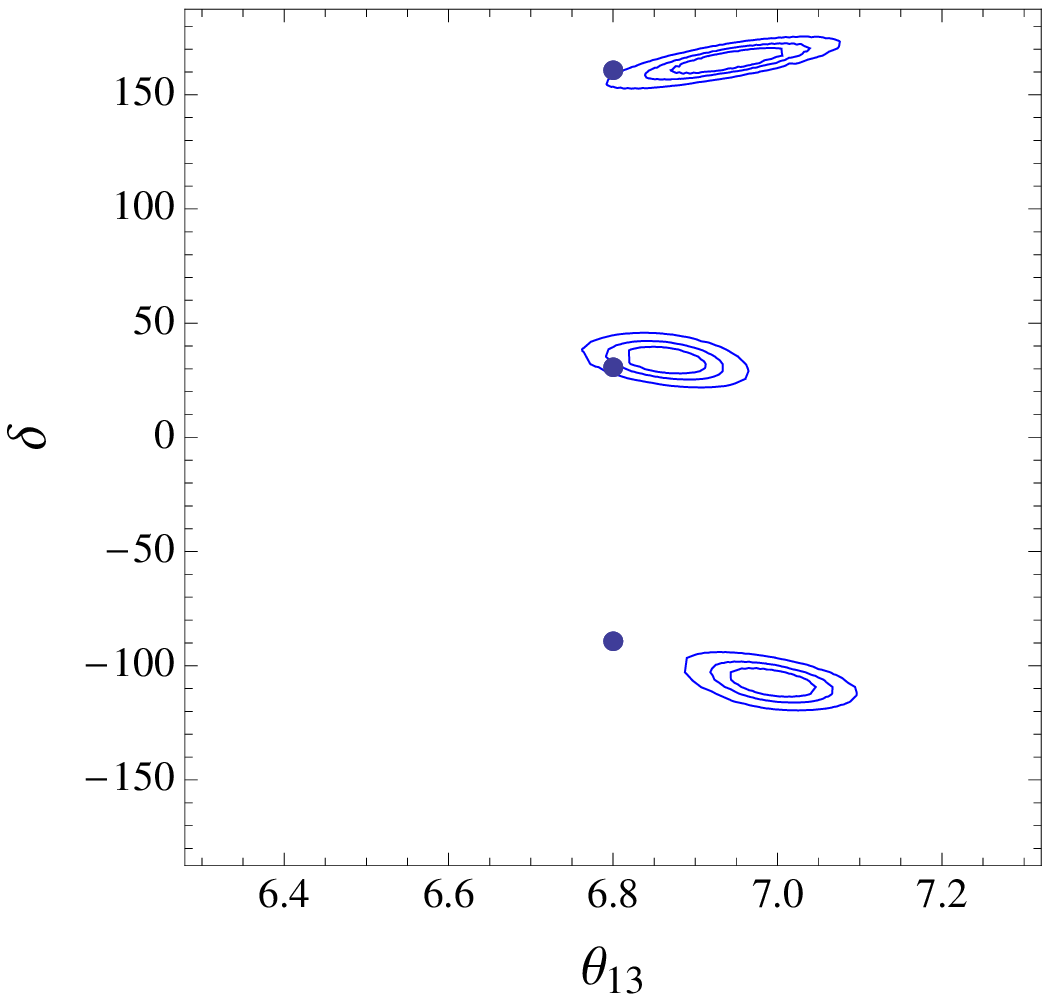} &
\hspace{-0.3cm} \epsfxsize6cm\epsffile{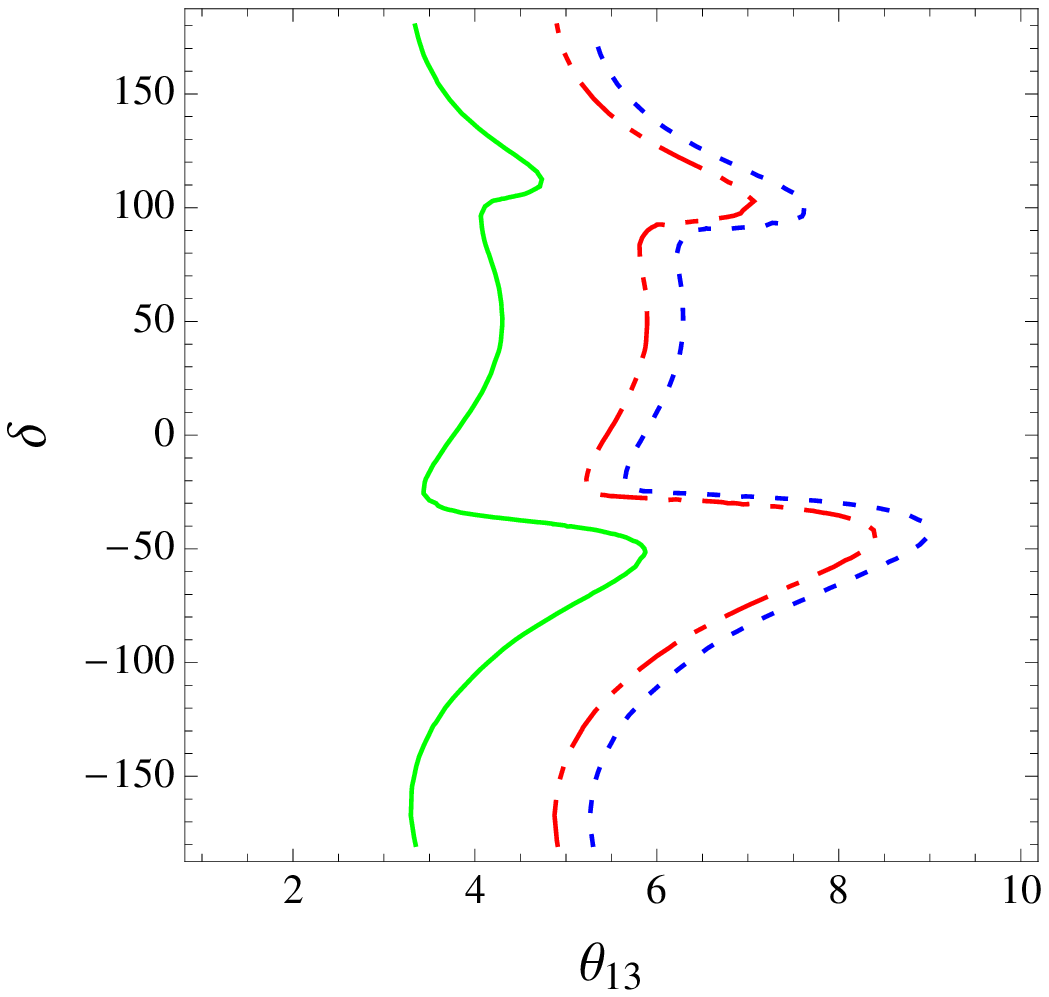}
\end{tabular}
\caption{\it Left: $\Delta \chi^2$ contours at 1, 2 and 3 $\sigma$ (2 dof's) of a fit of the golden muon theoretical distribution to
simulated data that include the $\tau$-contamination. 
The data have been produced for $\theta_{13} = 6.8^\circ$  (corresponding to the present best-fit value from the global fit to solar, atmospheric and LBL data, from Ref.~\cite{GonzalezGarcia:2010er}) and $\delta = 160^\circ, 30^\circ$ and $-90^\circ$. The dot represents the input value. 
Right: Test of the hypothesis that a simulation of the data that includes the effect of the $\tau$-contamination can be fitted with the golden muon theoretical distribution.
In the regions to the right of the contour lines, the hypothesis can be rejected at 1, 2 or 3$\sigma$ (from left to right), assuming the goodness-of-fit statistics follows the $\chi^2$ distribution with $n =8$ dof's. In both panels, data refer to the $L= 4000$ Km baseline.}
\label{fig:taucontbestfit}
\end{center}
\end{figure}

In Fig.~\ref{fig:taucontbestfit}(left) we show the result of a fit to the data in the ($\theta_{13},\delta$)-plane performed for the input value $\theta_{13} = 6.8^\circ$ (corresponding to the present estimate from the global fit to solar, atmospheric and LBL data from Ref.~\cite{GonzalezGarcia:2010er}) for three representative values of $\delta$, $\delta = 160^\circ, 30^\circ$ and $-90^\circ$. Data have been obtained for a 25 GeV Neutrino Factory, a 50 Kton MIND located at $L = 4000$ Km from the source, with $5 \times 10^{20}$ useful muon 
decays per year per baseline and 5 years of running time with each muon polarity. 
Events are binned in the reconstructed neutrino energy, with five bins of constant size $\Delta E_\nu = 5$ GeV. The input parameters of the simulation, in addition to $(\theta_{13},\delta)$, have been kept fixed to: $\Delta m^2_{21} = 7.6 \times 10^{-5}$ eV$^2$, $\Delta m^2_{32} = 2.5 \times 10^{-3}$ eV$^2$, $\theta_{12} = 33^\circ$ and $\theta_{23} = 42^\circ$ \cite{GonzalezGarcia:2010er}.  The data correspond to a MonteCarlo simulation that includes the golden muon sample and the $\tau$-contamination. On the other hand, the fit to data (in this section) is always performed using the theoretical distribution of golden muons (cross-sections and efficiencies taken from Fig.~\ref{fig:efficiencies}, right, are properly taken into account). 

It can be clearly seen that the best-fit point does not coincide with the input value (represented by the dot), independent of the particular choice of $\delta$. At the intermediate baseline (left panel), we see that  the input value lies generally at the border of the 3$\sigma$ contour of the fit, but for  the case of $\delta = -90^\circ$, when it lies even further away. 
Notice that the $\chi^2$ at the best-fit point is, in all cases, rather large: for $\delta = 30^\circ$ (the case in which the distance between the best-fit point and the input value is the smallest), we get for the normalized $\chi^2$ distribution $\chi^2_{min} \sim 3.63$ (for $n = 8$ dof's). These results imply that less than $0.01$\% of the time we would have got a fit to the data worse than what we have found, within the hypothesis that the data are distributed accordingly to the golden muons theoretical distribution. In practice, the hypothesis can be rejected at more than 3.9$\sigma$ for all of  the considered input pairs.


The result of this analysis can be generalized to different values of the true $\theta_{13}$. In Fig.~\ref{fig:taucontbestfit}(right) we present the result of a test of the hypothesis that a
 simulation of the data that includes the effect of the $\tau$-contamination can be fitted by a theoretical distribution including golden muons, only. 
The area to the right of the contour lines represent the region of the ($\theta_{13},\delta$) parameter space for which the hypothesis can be rejected at 1, 2 or 3$\sigma$ (from left to right), assuming the goodness-of-fit statistics follows the $\chi^2$ distribution with $n =8$ dof's. 
The hypothesis that the golden muon theoretical distribution can fit the data can be rejected at more than 3$\sigma$ for a true $\theta_{13} \geq 6^\circ$ for almost any value of $\delta$
(but for relatively small regions around $\delta = \pm 90^\circ$). In this part of the parameter space, clearly, we would be obliged to use a different distribution to fit the data. 

\begin{figure}[t!]
\vspace{-0.5cm}
\begin{center}
\begin{tabular}{cc}
\hspace{-0.3cm} \epsfxsize7cm\epsffile{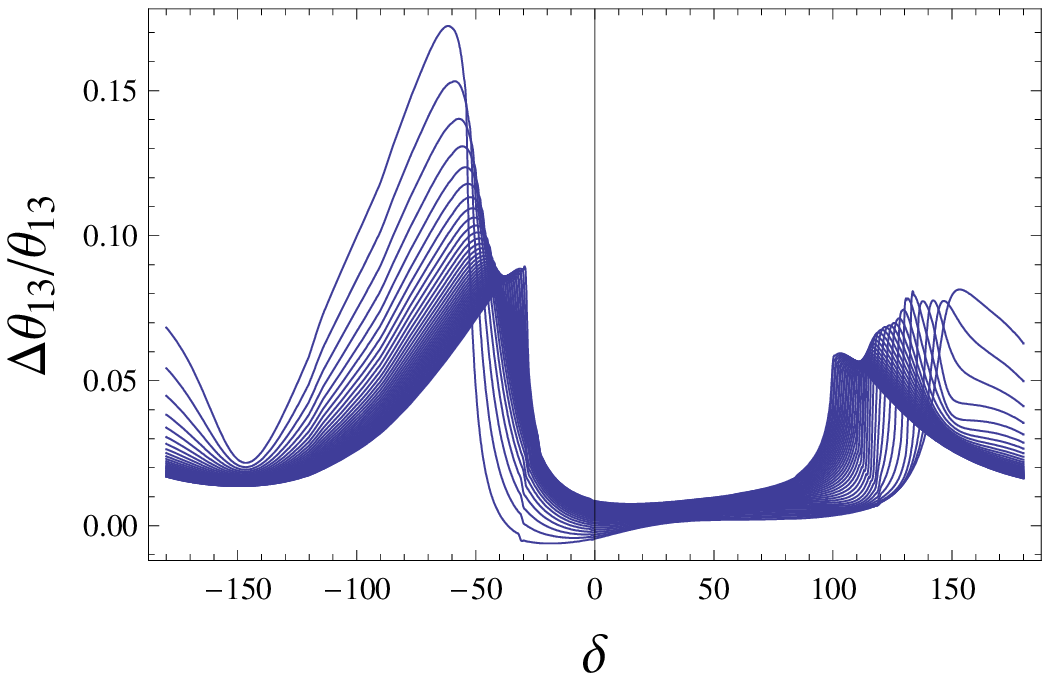} &
\hspace{-0.3cm} \epsfxsize7cm\epsffile{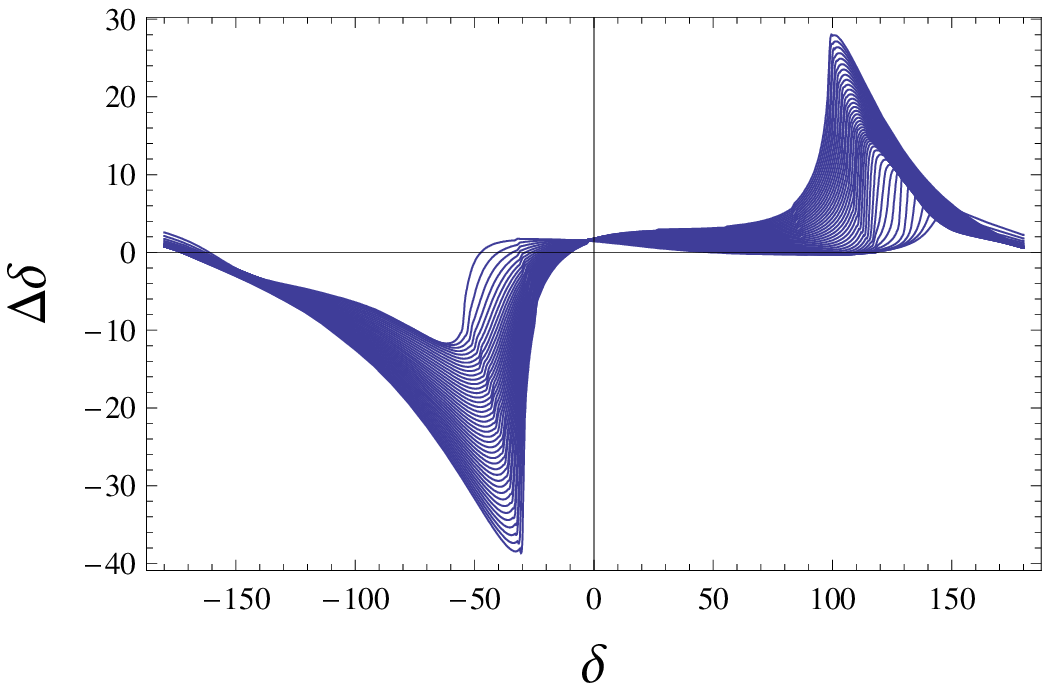}
\end{tabular}
\caption{\it  Left: the value of the relative shift in $\theta_{13}$ at the best-fit point; Right: the value of the absolute shift in $\delta$ at the best-fit point. 
In all cases, results are shown as a function of the true $\delta$. Different lines correspond to different true $\theta_{13}$ in the range $\theta_{13} \in [3^\circ,10^\circ]$. 
Data have been obtained at $L = 4000$ Km.}
\label{fig:shift}
\end{center}
\end{figure}

The scenario can, in a sense, be even worse when $\theta_{13}$ is smaller than $5^\circ$. In this part of the parameter space, we cannot exclude with high statistical significance the hypothesis that the data are statistically distributed accordingly to the golden muon theoretical distribution, although we would still get a poor fit to the data. Moreover,
 for input parameters in this region the best-fit point of the $\chi^2$ fit to the data is always located far from the input value. 
The distance between the best-fit point of the $\chi^2$ fit to the data obtained using the golden muon theoretical distribution and the input values ($\theta_{13},\delta$) is shown in 
Fig.~\ref{fig:shift} for several values of the true $\theta_{13}$ in the range $\theta_{13} \in [1^\circ,5^\circ]$, as a function of the true $\delta$. 
Since the considered range for $\sin^2 2 \theta_{13}$ runs over several order of magnitude, in Fig.~\ref{fig:shift}(left) we present the relative error in $\theta_{13}$ (i.e., 
the distance in $\theta_{13}$ between the best-fit point of the $\chi^2$ distribution and the input value normalized to the input value, $\Delta \theta_{13}/\theta_{13}$). 
On the other hand, since at present $\delta$ can assume any value in the range $[0,2 \pi[$, the distance in $\delta$ is shown in Fig.~\ref{fig:shift}(right) in terms of the absolute error, 
$\Delta \delta$.

As can be seen from Fig.~\ref{fig:shift}(left), the relative error on $\theta_{13}$ can be as large as 17\%. The maximal relative error is found in the region of negative $\delta$ and small 
$\theta_{13}$, $\theta_{13} \in [1^\circ,2^\circ]$ ($\sin^2 2 \theta_{13} \in [1,4] \times 10^{-3}$). The absolute error in the measurement of the CP-violating phase can also be huge: 
from Fig.~\ref{fig:shift}(right) we can see that $|\Delta \delta|$ can be as large as $40^\circ$.  The maximal error is found, again, in the region of negative $\delta$, 
but for relatively large $\theta_{13}$. For the same regions of the parameter space we have found that, combining the intermediate and the magic baseline data, the $\theta_{13}$ relative error and the $\delta$ absolute error can still be as large as 4\% and $12^\circ$, respectively. Clearly, such large errors prevent the high-energy Neutrino Factory setup from measuring the value of the mixing angle $\theta_{13}$ with precision at the level of the percent. If $\theta_{13}$ is smaller than $\theta_{13} \sim 5^\circ$, a wrong treatment of the $\tau$-contamination 
will prevent the use of the Neutrino Factory as a precision machine and will introduce a systematic error in the joint measurement of $\theta_{13}$ and $\delta$. 
 
It is worth noting that the $\tau$ contamination problem could also affect other observables, like the 
CP-discovery potential, that we define as the capability of a Neutrino Factory 
to measure a value of the CP phase $\delta$ different from the CP-conserving cases $\delta=0,\pm \pi$, at some confidence level\footnote{We thank Thomas Schwetz for rising this point.}. 
A loss in sensitivity could arise when the silver muons, properly taken into account,
add to the golden sample: in that case, due to the fact that the CP violating terms in the $\nu_e \to \nu_\mu$ and 
$\nu_e \to \nu_\tau$ probabilities have opposite signs, we would expect a suppression of the bulk of the events which depend on the CP phase $\delta$. We carefully checked that this is not the case, mainly because the silver statistics is irrelevant at very small $\theta_{13}$ when compared with the golden channel.

In this section we have shown at length that a wrong treatment of the $\tau$-contamination is extremely troublesome if the value of $\theta_{13}$ is larger than $1^\circ$ (as suggested by
present three-family oscillation fits) and prevents the use of the Neutrino Factory as a precision facility for large $\theta_{13}$.
Similar results on the impact of $\tau$-contamination in the atmospheric sector have been found in Ref.~\cite{indians}, where it was shown that the measurement of $\theta_{23}$ 
when $\tau$-contamination is taken into account has an error that is almost twice as large as when muons from $\tau$'s are correctly removed at 99\% CL for the input
pair $\theta_{23} = 42^\circ, \Delta m^2_{32} = 2.4 \times 10^{-3}$ eV$^2$.  This means that the precision on $\theta_{23}$ falls from 5\% to $\sim$10\%.
As a consequence, the $\tau$-contamination severely reduces the ability of the high-energy Neutrino Factory setup to discriminate a non-maximal $\theta_{23}$ 
from $\theta_{23} = 45^\circ$ and the capability of the Neutrino Factory to solve the octant degeneracy \cite{Meloni:2008bd}.  
On the other hand, the measurement of the atmospheric mass difference $\Delta m_{32}^2$ seems to be less affected, and it remains at the percent level.

Notice that one of the main difference between Ref.~\cite{indians} and our analysis, apart from the study of different channels and parameters,  is that in Ref.~\cite{indians} 
the signal is studied as a function of the final muon energy, whereas in this paper we analyse data as a function of the reconstructed neutrino energy. Although binning in the final muon energy allows the addition of the $\tau$-contamination as a signal source to the wrong-sign muons sample, as we have explained in Sect.~\ref{sec:taucont} this prevents an optimal use of the detector hadronic calorimetry that, in turn, translates into a worse treatment of the backgrounds. A detailed analysis of the $\tau$-contamination impact on the measurement of the atmospheric parameters $\theta_{23}$ and $\Delta m^2_{32}$ when data are analysed as a function of the reconstructed neutrino energy can be easily done, 
but it is beyond the scope of the present paper and it will be presented elsewhere. 

%
\section{The solution to the $\tau$-contamination problem}
\label{sec:solu}
%
%

In the previous sections we have discussed at length the problems related to fitting, using the theoretical golden muon distribution, simulated data that include the true golden muon 
sample and its corresponding $\tau$-contamination. We will now explain how the simulation is performed and how we can use the same procedure used to produce the data to 
actually solve the problem. 

Consider a $\nu_\tau$ of energy $E_{\nu_\tau}$, interacting in MIND and producing a wrong--sign $\tau$ of energy $E_\tau$ together with a hadronic jet of energy $E_h$.
The heavy lepton decays subsequently ($\sim$ 20\% of the times) into two neutrinos and a muon, with $E_\tau = E_\mu + E_{miss}$, where $E_{miss}$~is the 
missing energy carried away by the two neutrinos in the $\tau$-decay. We have, therefore, 
\begin{equation}
\label{eq:Enutau}
E_{\nu_\tau} = E_\tau + E_h = (E_\mu + E_{miss}) + E_h \, .
\end{equation}
Experimentally, we observe the secondary muon and a hadronic jet, a signal essentially indistinguishable from that of a wrong--sign muon from CC $\nu_\mu$ interactions.
However, in this latter case, the addition of the (primary) muon energy $E_\mu$ and of the hadronic jet energy $E_h$ results in the correct parent $\nu_\mu$ energy, 
$E_{\nu_\mu} = E^\mu + E_h$. On the other hand, in the former case the addition of the (secondary) muon energy $E_\mu$ and of the hadronic jet
energy $E_h$ results in the wrongly reconstructed fake neutrino energy $E_{fake}$. As can be immediately seen from eq.~(\ref{eq:Enutau}),
the relation between $E_{\nu_\tau}$~and $E_{fake}$~is simply:
\begin{equation}
E_{fake} = E_\mu + E_h = E_{\nu_\tau} - E_{miss} \, .
\end{equation}
The decays of the produced $\tau$ would result in muons with all the possible energies between $m_\mu$ (the case in which the neutrinos carry most of the energy of the $\tau$) and the 
$\tau$ energy (the case in which the neutrinos carry no energy). If we divide the continuous distribution of the $\tau$ three-body decay in discrete fake neutrino energy bins,
we find that for a monochromatic $\nu_\tau$ beam of energy $E_{\nu_\tau}$, the final muon will be assigned to a given fake neutrino energy bin of energy $E^\mu_j$ 
with probability $V_j(E_{\nu_\tau})$,  where $j = 1, \dots, N^\mu_{bin}$. 
The neutrino factory does not produce a monochromatic $\nu_\tau$ beam, however. Since the beam's flux is exactly calculable, we can compute the distribution of $\nu_\tau$
of a given energy $E_{\nu_\tau}$ and divide them into $\nu_\tau$ energy bins of energy $E^\tau_i$, where $i = 1, \dots, N^\tau_{bin}$.
The ensemble of the probability vectors $V_j (E^\tau_i)$, for $i$ and $j$ running over all the $\nu_\mu$ and $\nu_\tau$ energy bins, is represented by the migration matrix $M_{ij}$. 

\begin{figure}[t!]
\vspace{-0.5cm}
\begin{center}
\hspace{-0.3cm} \epsfxsize9cm\epsffile{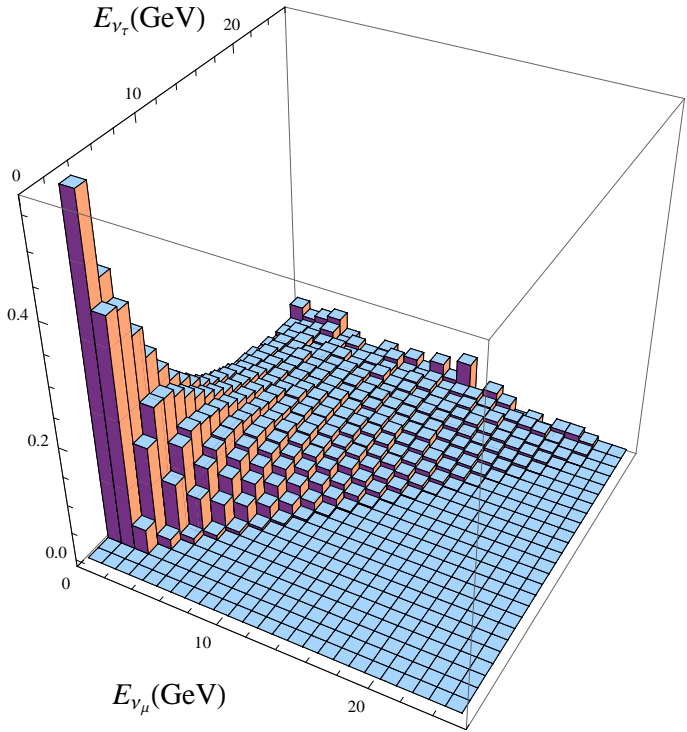}
\caption{\it The migration matrix $M_{ij}$.}
\label{fig:migmat}
\end{center}
\end{figure}
$M_{ij}$ is, of course, not measured by experimental data taking. However, knowing the NF neutrino flux, the oscillation probabilities, the differential $\nu_\tau N$ cross-section 
and the differential $\tau \to \mu$ decay width, we can compute statistically $M_{ij}$ through a MonteCarlo simulation of the events. This is the procedure we have used to simulate
data that include both the true golden muon signal and the corresponding $\tau$-contamination. The $M_{ij}$ migration matrix has been computed using the GENIE neutrino generator \cite{Andreopoulos:2009rq}  with  
$10^6$ simulated $\nu_\tau$'s per neutrino energy bin and 25 bins  in the range $E_{\nu_\tau} \in [0,25]$ GeV. The explicit form of the migration matrix $M_{ij}$ is depicted in Fig.~\ref{fig:migmat}. In this figure, we show the statistical distribution of the fraction of the events with $\nu_\tau$ of energy $E^\tau_i$ that produce a wrong-sign muon whose energy, 
combined with the hadronic energy $E_h$, will be erroneously assigned to the $\nu_\mu$ energy bin $E^\mu_j$.

\begin{figure}[t!]
\vspace{-0.5cm}
\begin{center}
\hspace{-0.3cm} \epsfxsize9cm\epsffile{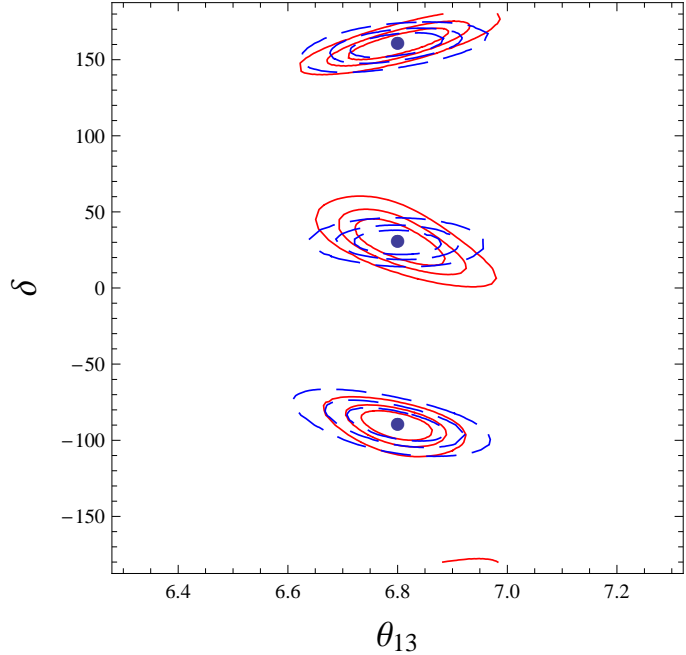}
\caption{\it $\Delta \chi^2$ contours at 1, 2 and 3 $\sigma$ (2 dof's) of a fit of the corrected muon theoretical distribution to simulated data. Dashed lines have been obtained
considering the golden muons only whereas the solid ones take into account the contribution of the silver 
events. 
The data have been produced for $\theta_{13} = 6.8^\circ$  (corresponding to the present best-fit value from the global fit to solar, atmospheric and LBL data, from Ref.~\cite{GonzalezGarcia:2010er}) and $\delta = 160^\circ, 30^\circ$ and $-90^\circ$. The dot represents the input value. 
Data refer to the $L= 4000$ Km baseline.}
\label{fig:problemsolvedbestfit}
\end{center}
\end{figure}

After having computed $M_{ij}$, the number of wrong-sign muons in a given neutrino energy bin is
\begin{equation}
\label{eq:ntot}
N_i = \sum_{i = 1, N_{bin}} \left [ N^\mu_i + \sum_{j = 1, N_{bin}} M_{ij} N^\tau_j \right ] 
\end{equation}
Once we know the theoretical distribution of the expected experimental muon sample, including both the true golden muon component and the corresponding $\tau$-contamination, we can use eq.(\ref{eq:ntot}) to fit the experimental data. The results of this fit for
the same input parameters as in Fig.~\ref{fig:taucontbestfit}(left) (i.e., $\theta_{13} = 6.8^\circ$ and $\delta = 160^\circ,30^\circ$ and $-90^\circ$ from top to bottom) are shown 
in Fig.~\ref{fig:problemsolvedbestfit}. As before, simulated data have been obtained for a 25 GeV Neutrino Factory, a 50 Kton MIND located at $L = 4000$ Km from the source, with $5 \times 10^{20}$ useful muon decay per year per baseline and 5 years of running time with each muon polarity. 
Events are binned in the reconstructed neutrino energy, with five bins of constant size $\Delta E_\nu = 5$ GeV. The other parameters used in the simulation are: $\Delta m^2_{21} = 7.6 \times 10^{-5}$ eV$^2$, $\Delta m^2_{32} = 2.5 \times 10^{-3}$ eV$^2$, $\theta_{12} = 33^\circ$ and $\theta_{23} = 42^\circ$ \cite{GonzalezGarcia:2010er}. 
As it can be seen, the $\tau$-contamination problem has been completely solved and the best fit
points coincide with the input points for all considered input pairs. The dashed lines have been obtained
considering the golden muons only whereas the solid ones take into account the contribution of the silver 
events. A part from a moderate $\delta$ dependence, which does not produce any relevant precision modification,
we cannot appreciate any major differences among the two sets of data. We also checked that this situation 
is common to other points in the $(\theta_13,\delta)$-plane and we safely conclude that the silver sample 
does not have any significant impact on the determination on the two unknowns 
(as also found for the CP discovery potential).

%
\section{Conclusions}
\label{sec:concl}

In this paper we have shown that:
\begin{itemize}
\item The ``tau contamination'' problem, ignored in early analyses of the Neutrino Factory sensitivity introduces, if not properly treated an intolerable systematic error, in particular for 
large $\theta_{13}$. In particular, for $\theta_{13} \geq 5^\circ$, the golden muon theoretical distribution is not able to fit the $\tau$-contaminated data. Data, in this case, 
can only be properly studied using the final muon energy distribution, thus reducing the capability of the MIND detector to handle the background and keeping it at an acceptable level.
For $\theta_{13} \in [1^\circ,5^\circ]$, the error in the joint measurement of $\theta_{13}$ and $\delta$ can be so large that it could actually prevent the use of the Neutrino 
Factory as a precision facility. 
\item This problem arises because recent analyses of the Neutrino Factory use muon neutrino energies above 1 GeV, while early analysis had an effective threshold of 10 GeV.
It was shown that, as a consequence of this, the CAD problem at high-energy Neutrino Factory setups was extremely severe. The inclusion of neutrino energy bins below 10 GeV
is very important to mitigate the CAD problem, and for this reason new (softer) kinematical cuts were applied to the golden muon sample in order to recover efficiency at low energy. 
\item Taus contaminate low energy bins and cannot be separated from the wrong sign muon signal by means of kinematical cuts. However, the use of a migration matrix allows to compute their contribution to the signal bin-by-bin. When such contribution is properly introduced in the fit, the large systematic error in the determination of  $\theta_{13}$~ and $\delta$~introduced otherwise is solved. Calculation of the migration matrix is straightforward since the relative weights of the different $\nu$ flavours are exactly known at the Neutrino Factory. 
\end{itemize}

%
%

%
\section*{Acknowledgments}
%
We gratefully acknowledge A. Cervera, P. Hernandez, O. Mena and H. Minakata for several interesting discussions. 
We warmly acknowledge the help of C. Andreopoulos with the GENIE neutrino event generator.  
This work was supported by the European Union under the European Commission
Framework Programme 07 Design Study "EURO$\nu$",  project 212372 and  by the spanish ministry for Science and Innovation under
the program �CUP� Consolider-Ingenio 2010, project CSD2008-0037.
A.D. acknowledges funding by the spanish ministry for Science and Innovation under the project FPA2009-09017and by the Comunidad Aut\'onoma de
Madrid through project HEPHACOS-CM (S2009ESP-1473).
D.M. was supported by the Deutsche Forschungs-gemeinschaft, contract WI 2639/2-1. 
%


%
%
\end{document}